\documentstyle[preprint,pra,aps,epsf,12pt]{revtex}
\begin{document}
\draft
\title{Protein design: A perspective from simple tractable models}
\author{E.I.Shakhnovich }
\address{Harvard University, Department of Chemistry and Chemical Biology\\
12 Oxford Street, Cambridge MA 02138}
\date{19 March 1998}
\maketitle
\newpage

{\bf We review the recent progress in computational approaches
to protein design which builds on advances in 
statistical-mechanical protein folding theory. In particular, 
we evaluate the degeneracy of 
the protein code (i.e. how many sequences
fold into a given conformation) and outline a simple
condition for ''designability``   in a protein model.
From this point of view we discuss several popular
protein models that were used for sequence
design by several authors.  We evaluate the 
strengths and weaknesses of
popular approaches based on stochastic optimization in sequence space
and discuss possible ways to improve them  to bring them closer
to experiment. We also discuss how sequence design
affects folding and point out to some features of proteins
that can be deigned ''in'' or designed ''out''}

\section{Introduction}

The protein folding problem has two aspects: 
``direct'' (i.e. folding)
and ``inverse'' (i.e. ``protein design''). The main issue 
of the ``direct'' PF problem is to understand 
the basic physical chemistry 
of how protein sequences determine their structure. The
long-range goal of these studies is to  
predict protein conformation
from sequence.  The direct protein folding  problem 
has received much attention recently and considerable progress
was achieved, in understanding the general principles
that govern folding of protein 
chains \cite{MKES,BRYN_REVIEW,FER_COSB,COSB}.
Using the language of  
bioinformatics one can define the folding
problem as mapping the space of 
sequences into the space of structures.

The ``inverse'' protein folding problem 
is how to find a sequence that
folds into and is stable in a 
given conformation at a given temperature. (see Fig.1). 
Again using  
the language of bioinformatics we can say that this corresponds to the 
mapping of space of structures to the space of sequences.

\marginpar{Fig.1}

It is clear that the two problems 
are closely related to each other:
better understanding of the principles of protein folding 
makes it possible to clarify which features of protein 
sequences are necessary (as well as sufficient) for their
stability and fast folding, i.e.
what makes protein a protein. Such understanding focuses
the attention of designers on emphasizing those crucial features
of folding sequences.

The experimental approaches to protein structure 
determination have been very successful providing
a wealth of structural information. 
While the growing flow of
genomic information makes the
development of theoretical approaches to predict
protein conformation even more desirable, 
there is an experimental
''shortcut'' of X-ray crystallography or NMR to the
solution of the ''direct'' PF problem.

The situation with design is very different. 
Most of the present experimental
approaches enjoyed only limited 
success providing
polypeptides which in most 
cases fold into compact but mostly
disordered conformation of 
molten-globule like species 
(see e.g. \cite{EX_DES1}). 
It is quite possible that limitations
in experimental design are due to 
relatively low synergism
between experiment and theory 
in that area.
An important success story based on  such synergism 
of theory and experiment is
given in \cite{MAYO_SCI} 
where theoretical analysis
has helped to guide the design 
effort which resulted in
a small protein that folded into 
predicted ''target'' conformation.
This work clearly 
demonstrates a crucial role of theory
in protein design. A  limitation
of the approach reported 
in \cite{MAYO_SCI} is that it requires
complete enumeration of 
sequence candidates - a problem
that explodes exponentially 
with chain length and thus limits
this valuable approach to relatively 
short lengths. The success 
and limitation
of the  work of Mayo and 
coworkers call for further 
refinement of theoretical approaches 
to protein design some of
which will be outlined in this review. 

It is important to note that the 
bottleneck in protein design is 
not on the synthetic side, but
rather in the fundamental problem that 
researchers generally do
not know which sequences
to synthesize. Since the number 
of possible sequences is enormous,
and the fraction of them that are 
able to fold into protein-like structures
is negligible (see below) the 
probability to ``hit'' a correct sequence
by chance is vanishingly low. Of course there exist
clever experimental approaches, like phage display \cite{Baker_NSB}
 which bias 
experimental sequence search towards better  candidates. 
However, in our view, convincing 
success in protein
design will come with  reliable 
theoretical approaches which 
will make it possible to find sequences that fold uniquely
into a desired conformation. Perhaps this goal
alone justifies all the effort that
has been put into protein folding theory
over last few years.

In this review I will discuss 
how recent advances in understanding
protein folding help us in the 
efforts to design protein sequences
and understand their natural evolution.

\section{Mapping Structures into Sequences: 
How many protein sequences are there?}

The computational approach to protein design 
aims to find sequences that fold to a given structure,
in a particular model. The fundamental question, is 
if there is any solution to this
problem (for a model of course, we know that there
is one for proteins) and if yes, how many solutions are there,
i.e. how many sequences can fold into a given conformation.
This question can be addressed only if we understand
what features should a folding sequence have.
Such understanding  builds on
recent developments in  protein folding theory 
which elucidated some of the
properties of folding sequences 
\cite{GSW,PNAS,SSK1,KT}.

According to thermodynamic hypothesis \cite{Anfinsen} 
 sequences that 
fold into a given structure have
lowest energy (potential of mean force) 
in that structure, compared to energies of 
decoys, i.e other conformations
for that same sequence. 
The  ''consistency principle'' due to Go \cite{GO1}
and ''principle of minimal frustrations'' (PMF) by
Bryngelson and Wolynes apparently posited 
that the necessary condition for protein stability 
and fast folding
is that the native state has energy that is
 much lower than energies of the bulk of
misfolded states (decoys). Speaking
modern language one can say that
PMF  is actually equivalent to the
requirement of 
large energy gap in protein-like models

The results of analytical microscopic 
theory of heteropolymer folding 
\cite{BIOCH,SGS,SHARAD1,PGT} as well
as numerical 
studies \cite{SSK1,PNAS,PRLF} in lattice model 
are  consistent with the PMF. More specifically, it was
shown that in order for a sequence
to fold into a given native 
structure, its energy
in that structure should fall
below a certain threshold
$E_c$. $E_c$ is the energy at which  
the density of states for decoys vanishes: 
at $E \ge E_c$ the 
density of states  is very 
high so that many decoys
belong to that energy 
range (see insert in Fig.2).  The probability
that there will be a decoy,
structurally unrelated to the native conformation
and  
having energy
$E<E_c$, has been estimated in the Appendix to \cite{SSK1} 
to be $exp((E-E_c)/T_c)$, where $T_c$ is the temperature
of thermodynamic freezing transition in random heteropolymer.
(The thermodynamic freezing transition is defined as temperature
at which entropy of a polymer vanishes \cite{BIOCH,GROS_BP}).
Therefore if a sequence folds into 
a given structure with energy 
$E$, the probability that there
will be structurally dissimilar
decoy having equal or lower energy 
 falls off exponentially and
 for sequences that fold into the target structure
with sufficiently low energy $E$ such that $E-E_c \gg T_c$, 
the target structure
will almost certainly be
a unique ground state conformation.

\marginpar{Fig.2}

Further studies showed that
pronounced  ''stability gap'' $E-E_c$
is also sufficient to provide fast 
folding for lattice model proteins
of considerable length 
(more than 100 monomers) \cite{PRLF,GROS_JCP}, 
consistent with the PMF \cite{B.W.}.

Therefore a possible search criterion for 
folding sequences is
large (many $kT_c$) stability  gap.
With that the issue  of how many sequences can
fold into a given conformation
(degeneracy of the protein code)
is reduced to the question of
how many sequences $\cal{N}(E)$ 
exist that have energy $E<E_c$ in a given structure:

\begin{equation}
\cal{N}(E)=\sum_{seq} \delta(H(seq,conf)-E)
\label{eq:N_E}
\end{equation}

Where $H(seq,conf)$ is energy of a particular sequence in the
target conformation. Delta means  that summation is
taken over all
sequences that have energy $E$
in the native conformation. A particular example 
which got much attention in the past \cite{MJ,KGS,PRL,MKES,SO2} is
when $H$ is a contact potential:
\begin{equation}
H(\{\sigma\},\{r\}=\sum_{i < j} (U(\sigma_i,\sigma_j))\Delta(r_i,r_j)
\label{eq:HAM}
\end{equation}
where N is the number of residues in the chain. The symbol 
$\sigma_i$ characterizes the type of monomer $i$ so that sequence
of monomers is defined as sequence of symbols
$\{\sigma\}$. There are 20 types of aminoacids so that $\sigma_i=1...20$.
The parameters $U(\sigma_i,\sigma_j)$ determine 
the magnitude of contact interaction
between monomers of type $\sigma_i$ and $\sigma_j$; 
several sets of such parameters
were published (\cite{MJ,KGS,Sippl,MS}). 
A simple approximation of conformation of a chain is residue
representation whereby a residue $i$ is assigned a one point location
variable $r_i$ (it can be a geometrical center of the 
side-chain or coordinate of its
$C_\alpha$ or $C_\beta$ atoms).

$\Delta(r_i,r_j)=1$ if residues i and j are 
in contact and 0 otherwise. For protein structures a reasonable
definition of a contact is  when distance 
between their $C\alpha/C_\beta$ atoms is less
than 6.5A (\cite{MJ}). For lattice model
proteins  definition of a contact is even simpler:
two aminoacids that are lattice, but not  sequence 
neighbors  are considered contacting.

$\cal{N}(E)$ in eq.(\ref{eq:N_E})
can be evaluated using  the  technique that  
represents Dirac delta-function in eq.(\ref{eq:N_E} via 
Fourier transform, expands appearing exponentials up to
the second order, sums over all sequences and re-exponentiates
the result. The final result of the calculation can be expressed
in terms of ''entropy'' in the sequence space:

\begin{equation}
S_{seq}(e)=\ln \cal{N}(E)=\log(m_{eff})-\frac{(E-E_{av})^2}
{2ND^2}
\label{eq:s_seq}
\end{equation}

$m_{eff}$ is the effective number of types of aminoacids:
\begin{equation}
m_{eff}=exp(-\sum_{i=1}^{20} p_i \ln p_i)
\label{eq:m_eff}
\end{equation}
(e.g. if all types of aminoacids are equally 
represented so that $p_i=1/20$ for any $i$ then 
$m_{eff}=20$. In the opposite case when, say $p_1=1$ and 
$p_i=0$ for any $i=2...20$ then $m_{eff}=1$ which makes 
clear sense since the latter situation corresponds to a homopolymer.)
$E_{av}$ is average (over all
conformations) energy of interactions, per aminoacid. and 
$D$ is the dispersion of interaction energies (per contact). 
$E_{av}$ is calculated as an average interaction energy 
over all {\em possible}
contacts; It depends on aminoacid composition but
not on details of the sequence. $D$ is dispersion
of contact energies also calculated over all {\em possible}
contacts. Calculation of these quantities does not require 
simulations or enumerations in conformational space. However,
certain geometrical properties which may restrict 
the types of possible contacts should be taken into account.
can be taken into account.
For example, for a cubic lattice an important property is
that only possible contacts are 
between units with odd and even positions along the chain.
This ''even-odd'' rule should be   taken into account
in estimate of $E_{av}$ and $D$ for cubic lattice model.

(The question of how many sequences fold into
a given structure was first addressed by Finkelstein,
Badretdinov and Gutin who postulated 
the distribution given in eq.(\ref{eq:s_seq}) \cite{FBG}).

According to 
the heteropolymer theory \cite{B.W.,BIOCH,NATUR,GROS_BP}
the density of states of 3-dimensional heteropolymer 
(the number of conformations
having energy in a given range) 
follows the Random Energy Model distribution:

\begin{equation}
W(E)=\gamma^N exp \Big(-\frac{(E-E_{av})^2}
{2N D^2} \Big)
\label{eq:REM}
\end{equation}

The energy at which the 
chain runs out of states (the boundary of the continuous
spectrum $E_c$  in the insert in Fig.2) is 
estimated from the condition 
$W(E) \sim 1$, i.e. 
\begin{equation}
E_c-E_{av}= N(2\ln \gamma)^{1/2}D
\label{eq:e_c}
\end{equation}

As explained above, a necessary 
condition that determines a folding sequence 
is that its energy in the
native state is $E < E_c$. Such 
sequences should exist, i.e. 
$S_{seq}(E<E_c)>0$.
It follows from \ref{eq:e_c} and  
\ref{eq:s_seq} that this
condition can be satisfied only when
\begin{equation}
m_{eff} > \gamma
\label{eq:main_c}
\end{equation}

Apparently,
there is another threshold
energy, $E_{lowest}$ such that there are no
sequences that have energy in the native state 
lower than $E_{lowest}$. A possible crude estimate of $E_{lowest}$
can be obtained  from the condition that at 
this energy the
system runs out of {\em sequences}. Mathematically this  
is equivalent to  
the condition 
$S_{seq}(E_{lowest})=0$. However it is quite possible 
that this is an overestimate and the actual boundary of
lowest possible energies in a sequence model may be higher than
estimated from the entropy condition below.

Therefore, the upper bound estimate of the 
maximal possible gap $E_{lowest}-E_c$ is
\begin{equation}
G_{max}=N\ln \frac{m_{eff}}{\gamma} (2D^2)^{1/2}
\label{eq:G_max}
\end{equation}

\marginpar{Fig.3}

A specific simple example to clarify the main concepts of
this analysis 
is presented  in Fig.3.  It shows
the energy spectra, or densities of states 
(log of the number of conformations having a given energy) 
for the designed (black bars) and a random sequence 
having the same composition (13B, 14W) (grey bars). 
Comparing this spectrum with the one presented schematically
in the insert in Fig.2 
one should keep in mind that for the  model 
that has only two kinds of aminoacids the spectrum
is apparently discrete because 
possible values of energy are determined by numbers of contacts
of different kinds which are obviously integer
(a straightforward generalization of heteropolymer
results to this discrete case is given in \cite{DIS_REM}).
However, the occupancy at each energy level (i.e. 
how many conformations
have that energy)
is different for different levels. Specifically, 
there may be energy levels that 
are highly populated i.e. a multitude of conformations 
have that energy. There exist also empty
low-energy levels which can be filled only for special sequences
(i.e. only special sequences can have such an ''unusually'' 
low energy in their native conformations). 
The designed sequence shown in Fig.3 
has absolute lowest possible, for the model, energy 
$E_N=E_{lowest}=-84$ in its unique native conformation.

It can be seen clearly in Fig.3
that the spectra for the random and the 
designed sequences 
differ only at the low energy part: 
at energies that are higher or equal than
-60 both random sequence and the designed one have almost identical
spectra, i.e. this part of the spectrum is sequence independent 
(quantities that are sequence 
independent are called self-averaging \cite{MP,NATUR,BRYN_REVIEW}). 
According to
the heteropolymer theory \cite{BIOCH,NATUR,B.W.,GROS_BP} 
the density of states is self-averaging
at energies $E_c$ and higher while the low-energy  part at $E<E_c$ 
is sequence specific. The low-energy non-self-averaging
part of the spectrum  represents an energetic fingerprint 
 of a sequence.

It follows that
for this model $E_c=-60$. 
Note also the concave shape at the
left wing of the spectrum for designed sequence which 
is a signature of a cooperative transition \cite{GO1}. 
The cooperativity of transition (e.g. its widths) is directly related
to the value of the relative gap $g=(E_N-E_c)/E_N$.
For this model
$m_{eff} \approx 2$. Only compact conformations are considered, therefore
$\gamma=103346^{1/26} \approx 1.7$. The relative gap  is $g=-0.33$.

\section{Lessons for design}

The statistical-mechanical analysis suggests a number
of lessons.

{\em Lesson1: The design problem may be easier than folding
problem}. In a protein-like model where $m_{eff} >\gamma$
there is an exponential in chain length  $N$ number of sequences that
have sufficiently large  energy gap $G \sim ND$ to
fold reliably into the target structure.
Unlike folding where a unique ground state solution is sought,
in design any sequence having sufficient 
(not necessarily the greatest
possible) energy gap \cite{GSW,PNAS} folds 
cooperatively into the target conformation
if the temperature is not too low, see \cite{GLASS}. 
While the number of folding sequences 
is large, the {\em fraction} of folding sequences
(i.e. the probability to pick up a cooperatively folding sequence 
from the ensemble of random sequences) is quite low. That makes the
design problem nontrivial.

{\em Lesson2: ''The number of types of aminoacids may be an important
factor that determines the designability of a protein model''} 

The models where the number
of types of aminoacids $m_{eff}$ is small
are ''undesignable''. This means that even the best sequences
designed for these models have energy in the native state
higher than $E_c$, i.e. decoys with energy
lower or equal to the energy of the designed sequences
in the native state are present in such models. 
Apparently no folding is possible in this case
since the native structure is not unique.
An example of such undesignable model is
the so-called HP model \cite{WAGER}.

{\em Lesson 3: ''Stiffer'' chains provide greater energy gaps
and therefore are more designable}
The fundamental relation for a designable
model, the condition presented in (\ref{eq:main_c}) 
can be enforced either 
by increasing the number of aminoacid types or by decreasing
$\gamma$ i.e. by decreasing the 
number of conformations (per monomer). There is a number of ways
to decrease $\gamma$: formation of secondary structure,
forcing the conformational ensemble of a chain 
to the set of compact conformations  
(by introducing additional non-specific
attraction, Fig.3), biasing the conformations
to carry certain structural features (like in threading).
The example given in Fig.3 shows that even the 
''two-letter'' model
may  sometimes have non-degenerate native state
(but very small gap)
if its configurational space is 
restricted to only compact conformations. When full ensemble
is considered  the ground state of HP sequences become multiple
degenerate \cite{OTOOL2,PNAS,WAGER}. Apparently the number
of all conformations (per monomer)  $\gamma_{all}$ 
is greater than the number
of compact conformations $\gamma_{compact}$ so that
the condition (\ref{eq:main_c}) is violated for
the HP model when all conformations are considered.
On the other hand the ''two-letter'' models
 that are restricted to maximally compact conformations only 
are 
 just ''on the borderline'' of the validity
of the condition (\ref{eq:main_c}).

{\em Lesson 4:
 Protein design
 for most 3-dimensional models does not require  
''designing out'' the decoys; 2-dimensional models
behave very differently and require more complicated design that
may require ''designing out'' the decoys}. 

 The key to successful protein design 
is to find sequences that have low energy of the native state
without optimizing decoys at the same time. This factor 
increases the energy
gap or, equivalently, increases the thermal probability
to be in the native state (see below). To this end the 
''ruggedness'' of the conformational space of 3-dimensional 
random heteropolymer (as exemplified by the 
equivalence between heteropolymers
and the Random Energy Model (REM) \cite{B.W.,BIOCH,SGS})
plays a key role. 
According to the REM, most low-energy decoys
are structurally different from the native state
(except the ones that represent small fluctuations
around the native conformation - the native state ensemble).
To this end optimization of the 
native conformation energy
(i.e. making the native contacts stronger) does not
affect the low-energy structurally dissimilar decoys
(see Fig.3). 
That makes the designing ''in'' on the background
of decoys that are unaffected by sequence selection
efficient to 
increases the gap. 
We should emphasize that this
is true only for 3-dimensional models; in
two dimensions the optimization of the native
states gives rise to optimization of numerous
partly folded low-energy decoys making the native state 
unstable (in contrast to the 3D case where partly folded
decoys have high energy). The physical reason for such dramatic 
dependence on space dimensionality,
is  given
in \cite{GROS_BOOK,LOCAL} 
(especially see appendix to \cite{LOCAL}):
In 3-dimensional compact chains non-local contacts dominate while
in 2-dimensional chains local contacts play are dominant.

It was pointed out by several authors \cite{JPHYS,GROS_PRL,GOLDSTEIN}
that  some special 3-dimensional target conformations
(crumpled globules \cite{TOPCONSTR}) may be as ''undesignable''
by simple methods as two-dimensional models, for the
same reason - prevalence of local contacts.

\section{Stochastic Optimization in sequence space: 
Simple model solution for the design problem.}

The major lesson from the statistical mechanical theory
is that many  solutions of the design problem exist. A crucial
question of practical importance is how to find such solutions. To this end
a number of approaches, (reviewed in this chapter) 
of various complexity and efficiency 
have been suggested. 

It is clear that all what is needed
for successful design 
is to find a sequence $\{\sigma_i\}$ that has high 
thermal probability to be in the
native state:

\begin{equation}
P(T)=\frac{e^{{-\frac{H(\{\sigma_i\}, \{r_i^0\}}{k_bT}}}}{Z(\{\sigma_i\})}
\label{eq:P_T}
\end{equation}

Where the native state is characterized by the set of coordinates
of its residues $\{r_i^0\}$, $H$ is the energy of a given sequence
in a given conformation (cf(\ref{eq:HAM})). $Z$ is a partition function of the
chain
\begin{equation}
Z(\{\sigma_i\})=\sum_{r_i} e^{-{\frac{H(\{\sigma_i\}, \{r_i\}}{k_bT}}}
\label{eq:Z}
\end{equation}
where summation is taken over all conformations of the chain
$\{r_i\}$. T is temperature and $k_b$ is Boltzmann constant.

As presented      by eqs.(\ref{eq:P_T},\ref{eq:Z})  
the problem of design is of
great complexity since it involves search in both
conformational and sequence spaces. (The search in conformational
space is needed to determine the partition function). In other
words the ''exact'' solution of the design problem 
that includes exhaustive searches in conformational
and sequence spaces
would require
$(m_{eff}\gamma)^N$ ''trials'' - a prohibitive number
for any model of practical interest.

This calls for development of approximations that would allow 
to avoid exhaustive search both in sequence space
and in conformational space. A simplest  approach of this kind 
was
proposed in 1993 in \cite{PNAS}. It is based 
on the following ideas:

i) The optimization of stability is equivalent, in a simplest case, 
to the maximization
of the energy gap $g$ defined above (see Fig.1 of \cite{PNAS}
for a qualitative explanation of this fact). The boundary of the
continuous spectrum $E_c$ is a self-averaging quantity,
i.e. it depends on aminoacid composition
only while the lower part of the spectrum
$E<E_c$ is highly sequence specific. This conjecture
from heteropolymer statistical mechanics 
was shown to be correct for simple exact models, such as the
one shown in  Fig.3. 
It follows that the desired design results can be
obtained by selection of sequences that have low energy
in the target conformation at a given aminoacid composition.
It is clear that this statement is equivalent to the
assumption that the partition function 
$Z$ (more precisely contribution to $Z$ from non-native-like 
decoys) 
in the eq.(\ref{eq:P_T}) depends 
primarily on aminoacid composition rather than on sequence.
The analysis using  the Random Energy Model approximation 
suggests that 
this conjecture is valid at high 
enough temperature $T>T_c$
where $T_c$ is temperature of the 
''freezing'' \cite{B.W.,BIOCH,GROS_BP}
transition in a random heteropolymer 
having the same aminoacid composition.
 A lucid discussion of this point and further details 
can be found in \cite{GROS_BP}.

The gap optimization in  sequence space can be achieved by any stochastic
algorithm. In the case of sequence design  
the energy landscape in sequence space
is ''smooth'' \cite{PNAS,PE}, 
so that there is no complicated
search problem. Therefore a simple Monte-Carlo 
algorithm would suffice \cite{PNAS,PRLF,GROS_IMP,GROS_JCP}.

An {\em experimentum crucis} to test the 
statistical-mechanical approach to sequence design
is to pick an arbitrary conformation and design a sequence
that is expected to fold into that conformation.
A proof of concept
for a design method is 
an actual folding simulation of a designed sequence, 
starting from an arbitrary random coil
conformation. If the designed sequence 
converges to the target conformation 
and never encounters grossly misfolded conformations
with energy lower than the target conformation
then they may be stable in the target state,
and the design is successful.

This program has been carried out in \cite{PNAS,PRLF} 
where
random
mutations preserving the aminoacid composition
(monomer swaps) were introduced under Metropolis control 
with certain ''selective'' temperature $T_{sel}$. The model studied
in \cite{PNAS} is the same as shown in Fig.3. 
 Strong attraction  
between any pair of aminoacids  
shifted the conformational
ensemble in folding simulations 
towards compact states. 
The designed sequences were shown
to fold into the target (native) conformation which in all 
cases turned out to be the non-degenerate global energy minimum.

An attempt to carry out a rigorous test of
design for longer sequences (48-mers) in the HP model 
without introducing strong overall attraction was not successful: 
In that case the native conformation
was always  multiple degenerate. The  
 non-compact decoys often had lower energy
than the target conformation. These results
are consistent with earlier prediction \cite{PRLF} and the
presented statistical-mechanical analysis.

Therefore the two-aminoacid type model design cannot
be successfully extended to longer chains because 
of the requirement to restrict the conformational ensemble
by compact conformations only (see Fig.3).
Introduction of non-specific additional attraction 
to bias the conformational ensemble 
towards compact conformations
dramatically slows down folding making it infeasible to fold 
longer
chains \cite{BURSTCOLL,GLASS,WILBUR}. 
Thus the range of lengths that can be studied
using  the two-aminoacid type model is very limited. 
Such limitation may give rise to some  small-size artifacts.

An obvious solution of this problem 
is to use a greater number  of
kinds of aminoacids than just two.  This was done 
in  \cite{PRLF} where 
20 types of aminoacids and Myazawa-Jernigan
interaction potentials \cite{MJ} were used. 
The design-folding 
program was carried out for 20-aminoacid type 
model proteins on a cubic lattice
(with fixed composition corresponding to an ''average''
aminoacid composition in proteins). The designed sequences of 80-mers
folded fast and were stable in their target conformation;
No conformations with energy lower than the energy of the 
target conformation (for the designed sequence) were encountered.
These results provided, for the studied model,  
an important proof that design approach
based on statistical-mechanical theory of
protein folding is feasible 
and is basically correct, for the right model.  

 A somewhat different interesting 
approach to design was proposed by
Grosberg and coworkers \cite{GROS_IMP,GROS_JCP}. This approach is
based on the idea of pre-biological evolution by ''imprinting'', 
according to which first macromolecules could have evolved 
as a result of polymerisation of equilibrated monomers
which could have interacted with substrates at pre-polymerisation stage.
The ''imprinting'' design procedure also uses the MC annealing
protocol but in the system of disconnected aminoacids. After that
the chain is threaded through the ''annealed'' configuration
of monomers on the lattice, thus creating a sequence. The advantage
of this method compared to the design procedure
proposed earlier in \cite{PNAS,PE} is that it can be (in principle) 
experimentally realized
in an abiotic system. A disadvantage is that sequences obtained 
by ''imprinting'' are considerably less stable in their native conformation
and sometimes they may even 
not have the target conformation as global 
energy minimum. The reason is that sequence design uses the energy function
in which nearest neighbors in sequence do not interact (their interaction
adds a constant to energy of each conformation and therefore it is irrelevant).
The imprinting method does not take this factor into account, therefore 
when a chain is threaded through the annealed system of monomers 
it will often connect strongly interacting nearest neighbors, making them covalently
bound and therefore losing their strong attraction for stability
of the native state. Despite of that difficulty
it was demonstrated that the sequences obtained as a result of imprinting
procedure are often able to fold into their native  
conformation corresponding to global 
energy minimum \cite{GROS_IMP,GROS_JCP}.

 Several authors proposed other, than MC optimization techniques
to search sequence space \cite{JONES_DES,KD}. 
In our opinion, the MC search in sequence space is as efficient as
other optimization algorithms (because the landscape is smooth
and multitude of solutions exist).  However, the MC approach is advantageous
because it converges to the canonical 
distribution and hence
its results can be rationalized from 
the statistical mechanical perspective.

This interesting analogy between the statistics in sequence
space and several statistical-mechanical models 
was noted in \cite{PNAS,PE,PRLF,SW}.  The Hamiltonian
for sequence design eq.(\ref{eq:HAM}) (where the coordinates
are quenched but the aminoacid identity variables
$\sigma$ are allowed to vary) is analogous to 
the Hamiltonian of the Ising model
if there are only two types of aminoacids and 
to the Potts model if there are many types of
aminoacids.
It was pointed out in \cite{PNAS,PE}
that the MC design procedure converges to the canonical distribution
in sequence space. Therefore the statistics of sequences become
analogous to the statistics of ''spin configurations'' in
the equivalent statistical-mechanical 
models as it follows the same Boltzmann law. 
This analogy is explained in more detail in \cite{PE}
where the one-to-one correspondence between statistical characteristics of
sequence design and Ising model are listed in the Table 1. 
(Two-aminoacid type sequences were considered
in \cite{PE} but the results are trivially generalizable to
the multi-aminoacid type models). 

Of those  analogies probably the most important
one is the relation between entropy in  statistical-mechanical models, and
''degeneracy'' of the protein code. This analogy allows 
us to calculate 
$\cal{N}(E)$ directly from the MC
sequence design simulations. 
The idea of the calculation is based 
on the thermodynamic equation that
relates the entropy at a given temperature T with 
average energy at the same temperature via:
\begin{equation}
S(T)-S(\infty)=\frac{E(T)}{T}-\int_T^\infty \frac{E(t)}{t^2}dt
\label{eq:S_T}
\end{equation}
with $S(\infty)$ being entropy of a system at infinite temperature.
In our case of sequence design, 
the selective temperature, at which MC design 
procedure in sequence space is carried out, 
plays the role of temperature 
in eq.(\ref{eq:S_T}). $S(\infty)$ corresponds 
to random sequences without a bias towards
any particular structure.
$S(\infty)=N \ln m_{eff}$. The results of the calculation
are shown in Fig.1 for several proteins with the energy function
approximation given by eq.(\ref{eq:HAM}) (the sequence 
design simulations for each protein in Fig.1 were 
carried out keeping the aminoacid composition fixed 
and equal to the aminoacid composition of 
native sequence for each protein see \cite{PNAS,PE})). 
(The related results were presented in a
recent publication \cite{SW}). 
The solid line in
Fig.1 shows a theoretical estimate given by 
the eq.({\ref{eq:s_seq}). It is quite clear
that the theoretical estimate is in excellent 
agreement with the simulation
results. Further,
it is clear from Fig.2 that sequence entropy,
is approximately the same for 
all studied proteins (of course different
sequences fold into different 
protein structures; it is the {\em number}
of sequences that is invariant for different proteins). 
Such invariance
is understandable since in this approximation
the difference 
in energy functions eq.(\ref{eq:HAM}) between 
proteins are due the average coordination number
of their aminoacids and the connectivity, 
i.e which of the spatially
proximal aminoacids 
are sequence neighbors.
While these factors are crucial in determining which sequences
actually fold into a given conformation, they are not too specific
to give rise to pronounced differences in ''designability''.
This result of the analysis of the model with 20 types
of aminoacids can be compared with   the ''designability principle''
suggested by Finkelstein and co-authors \cite{FGB} and further
addressed by Tang and co-authors \cite{Tang}. 
The analysis presented in Fig.2 differs from that
of Finkelstein et al  that we did not impose energetic
penalties on certain structural features such as turns etc
while these factors were assumed to be important 
in \cite{FGB}. On the other
hand the arguments presented in \cite{FGB} 
are the phenomenological ones
that assume a certain form of density of 
states for a particular structure;
the justification of such assumptions based on a more 
microscopic model will be very interesting to obtain.

Tang and co-authors used a standard 27-mer models \cite{CHPH} 
with the form of energy function similar to eq.(\ref{eq:HAM}).
These authors  carried out exhaustive enumeration
of all compact conformations and all ''two-letter'' sequences.
The ''designability'' of a structure 
was defined in \cite{Tang} as the number
of sequences that have this structure 
as a unique energy minimum among
all compact conformations. Interestingly 
Tang et al report that  certain  structures
of compact 27-mers are more ''designable'' than 
others in their model.
Further they infer that the designable 
structures feature protein-like
properties such as secondary structure. 

It follows from the present analysis that the issue
of ''designability'' may be indeed important 
for the models that feature two kinds of aminoacids 
because some structures
can accommodate their ''best'' (lowest energy) sequences with slightly 
lower energies than  other structures can accommodate their ''best''
sequences. 
In the situation
when there is no significant gap, 
this small energy difference
between different structures
matters a lot: a more designable
structure can accommodate their sequences with 
energy slightly lower than $E_c$ while less
designable ones may have $E_{lowest}$ 
that is close or above $E_c$.
These factors can be clearly seen in Fig.3. For the structure
shown there the sequences with 
lowest possible energy $E_{lowest}=-84$ exist.
The lower the energy of the native state 
is the lower the probability that a decoy 
having the same energy will be found (see above and \cite{SSK1,DIS_REM}). 
Correspondingly
there may be many sequences that have the structure shown
in Fig.3 as their {\em unique} ground state, i.e. this structure
may be highly designable. It is clear that the designability
of this structure
is due to the special pattern of bonds on the lattice which
makes it possible to find a sequence that features complete separation
between beads of opposite kind (sequence neighbors
do not interact). However, there are many structures
that do not have such an ''ideal'' pattern of bonds so that
even their ''best'' sequences still have at least one contact
between aminoacids of opposite
kind. For them $E_{lowest}=-82$. 
For those sequences the gap is smaller
and therefore they are less designable
than the structure shown in Fig.3.
  This  
is  consistent with the observation 
of Tang and coworkers
that more designable structures deliver 
greater energy gaps \cite{Tang}.

This analysis implies that the pronounced difference
in designability exist for the models 
where even the maximal possible gaps are small (i.e $m_{eff} \approx \gamma$). 
In that case
every favorable contact matters a lot so that differences between
structures
 (patterns of bonds on the lattice) which allow 
to gain or lose an extra favorable contact may make a 
significant impact on
designability.    In 
many aminoacid kinds 3-dimensional
models where sequences can have energy in a target conformation
that is considerably below $E_c$ (i.e. $m_{eff} > \gamma$)
all structures may be highly designable. Therefore
it is important
to extend the study of \cite{Tang} to multi-aminoacid type model. 
However, such extension is a difficult one: It is computationally very costly
to enumerate the multi-letter sequences exhaustively as it was done
for two-letter sequences by Tang and coworkers 
\cite{Tang}. The MC simulations in sequence space 
may be a reasonable alternative to
exhaustive enumeration of sequences. The results 
presented in Fig.2 
show no visible differences in 
designability for a few protein 
structures which were used
for the analysis. 

An
important caveat of the MC sequence analysis should be mentioned 
here. The estimate of the number of sequences in eq.(\ref{eq:S_T}) is
based on the thermodynamic analogy which is not precise enough
to take into account sub-dominant (in $N$) contribution to entropy in sequence space. 
Therefore, though the major, exponential in chain 
length, contribution to the number of sequences 
that fold into a given structure 
(corresponding to the linear in $N$ contribution to sequence entropy),
is the same for different proteins, there may be sub-dominant
(less than exponential in chain length) contributions
which may give rise to some differences in designability. Whether
this is so and if yes, whether this is important for
our understanding of protein evolution is a matter
of future research.

The approach to the design which uses MC simulation in sequence
space with fixed aminoacid composition \cite{PNAS,PE,GROS_JCP} 
is  simple,
computationally very efficient and is non-heuristic one (i.e. it is
not limited to any particular model of a protein). Hence its appeal.

However, it has certain disadvantages most important 
of which are:

a) Keeping the aminoacid composition fixed 
eliminates the possibility to find an optimal (for folding and stability)
aminoacid composition.

b) The assumption of sequence independence 
of the partition function in eq.(\ref{eq:P_T})
(more precisely the contribution to it from non-native decoys) 
follows from the mean-field heteropolymer theory \cite{BIOCH,GROS_BP}. 
However, this assumption is valid 
only at high temperature. Furthermore, the deviations 
from the mean-field predictions need to be examined. 

c) The lack of reference to the temperature at which
sequence is expected to fold. Indeed, in the full 
design problem  sequence space 
optimization of $P(T)$ in the eq.(\ref{eq:P_T})
both the numerator and denominator
depend on temperature and it is possible that at different temperatures
different factors become important to optimize.

Those limitations were partially overcome in a number of subsequent
publications \cite{LOCAL,DK,MMES,SENO}. 

The first limitation (constant aminoacid composition)
was overcome in \cite{LOCAL,FD1} where 
the quantity $Z=(E_N-E_{av})/D$ 
(the so-called
$Z$-score, \cite{EISENBERG}) 
was optimized in sequence space.

Optimization of the Z-score 
instead of native energy 
fixed one of 
problems  of the simple approach \cite{PNAS,PE} 
- convergence to homopolymeric sequences
unless the aminoacid composition is 
constrained. As a result, the  design 
based on optimization
of the $Z$-score was able to find 
also optimal composition which
provided the best value of the gap.

A number of recent papers 
\cite{DK,MMES,SENO} addressed the second problem,
attempting to better estimate the partition function $Z$ 
than simply assuming it to be sequence-independent. 
In general this problem
is very complicated since an 
exact solution would require enumeration
of conformations after each 
mutation (to evaluate $Z$ for the new
sequence) which makes it 
computationally very difficult for small
chains and totally prohibitive 
for longer chains of realistic length.

The paper \cite{SENO} attempted to 
optimize directly $P(T)$ in eq.(\ref{eq:P_T}) 
using dual Monte-Carlo: in sequence
and conformational space (chain 
growth algorithm was applied
for conformational space simulation). 
This approach requires considerable computational
effort in order to reach 
Boltzmann distribution to provide a correct estimate of 
the partition function $Z$. 
Even for shorter chains such equilibration would require
more than $10^5$ MC steps and this number grows fast
with chain length \cite{LENGTH} making the 
interesting approach proposed by Seno et al \cite{SENO} very 
demanding computationally. 
The apparent advantage of this approach 
is that it contains direct reference to folding temperature
and is rigorous.
The disadvantage is that it is computationally very demanding 
for chains
if realistic lengths.

Deutsch and Kurosky (DK) attempted to estimate the 
partition function in high-temperature approximation
taking into account the first cumulant only by presenting
the partition function $Z$ in the simplest form:

\begin{equation}
F_s=-T \ln Z= =\sum_{1 \leq i < j \leq N} (U(\sigma_i,\sigma_j)) < \Delta(r_i,r_j >
\label{eq:DK}
\end{equation}
where the $<>$ denote unbiased averaging over all conformations.

It is quite clear that for compact chains the approach of
DK is basically equivalent to the earlier approach in \cite{PNAS} 
that
assumed sequence independence of the partition function.
Indeed in globular polymers the $< \Delta_{ij} >$
(which has the physical meaning of the probability
of a contact between monomers $i$ and $j$ in the
full ensemble of conformations) does not depend on
$i$ and $j$ except when these monomers are close
to each other along the chain \cite{GROS_BOOK,CONSTR}.
It is clear that setting  $< \Delta_{ij} >=const$
in eq.(\ref{eq:DK}) results in sequence independence
of the partition function. In apparent contradiction with the
above arguments DK reported a considerable improvement 
(for the 2-letter HP model) over the results of the 
previous approach \cite{PNAS}. 

It is possible that the 
improvement over the simplest design 
reported in \cite{DK} 
is due to the special
property of the cubic lattice 
that excludes the contacts for which $j-j$ is even. In other
words {\em on a cubic lattice}  $< \Delta_{ij} >= \approx const$
when $i-j$ is odd and is 0 otherwise.
The design in \cite{DK} took advantage of this property
of the cubic lattice providing proper distribution
of H and P monomers over even or odd sites. 

It is also worth mentioning that both Seno et al and DK
used the HP model to test the results of their
design procedures In both cases the methodologies are  not limited
technically to the HP model. 
As was explained before, the HP model is problematic
to study design and folding. 
For the two-letter model on the square lattice 
(as well as  on the cubic lattice with average attraction between monomers)
$m_{eff} \approx \gamma$, i.e. it is on the verge of failure. That
makes the design results for the HP model unstable
and heavily dependent on the details 
of a model such as lattice type, chain length,
''even-odd'' contacts, details of the composition etc. 
It is quite possible that some improvements of the design methods
over the simplest one suggested in \cite{PNAS} 
actually solve the problems specific to the
gapless HP model. Those problems 
may not exist in more realistic
multiple-letter models, where any reasonably compact 
structure is designable even within the simplest
algorithm of \cite{PNAS}.

To this end it would be desirable to apply interesting design 
methods proposed by DK and Seno {\em et al}
to
20 aminoacid types model and compare folding 
rates and stability of  sequences designed using various procedures.

Morrissey and Shakhnovich (MS) \cite{MMES} proposed a new design procedure
which seeks sequences having high probability to be in their 
native state at a given temperature $T$, $P(T)$. 
This procedure also employs MC in sequence space; 
however the partition function of the chain $Z$ entering the expression
for
$P(T)$ in eq.(\ref{eq:P_T})
is estimated using the cumulant expansion approximation. This eliminates the
need to run simulations in conformational space after each mutation 
to estimate the partition function 
\cite{SENO} and thus dramatically 
increases  the computational efficiency. 

This design procedure was carried out 
for 20-letter model proteins of various sizes (36-mers and 64-mers) 
on a cubic lattice and turned out 
to be quite efficient yielding sequences that
are stable at a selected temperature. 
Two interesting and unexpected results
emerged from this study: 
First, the {\em folding transition temperature}
for designed sequences turned out to be highly correlated with the input temperature
at which designed sequences were stable in their native conformations.

Second, the temperature at which 
folding {\em rate} was the fastest, appeared to be very close
to the stability temperature $T$ which was input in the
algorithm.  This reflects an important feature  of proteins
that optimum of their folding kinetics is achieved at the
conditions when their native state is not extremely stable - 
a finding fully consistent with the well-known marginal stability
of natural proteins. The reason for such relation between thermodynamics
and kinetics is partly given in a simple theory
of folding kinetics
presented in \cite{GLASS}. 

The observed correlation between folding rate and folding temperature 
generates an interesting prediction
that proteins from thermophylic organisms should fold very slow
at normal temperature (around $300K$) at which folding of mesophilic proteins
is fast. This prediction is partly supported by the observation
that  some thermophylic 
proteins (e.g. ribonucleotide Reductase from $Thermus X-1$ 
\cite{RNR}) are most active at high temperature (about $90C$)
and they retain only marginal activity at room temperature. The implicit
assumption made here 
is that enzymatic activity correlates with foldability. The validity
of this assumption requires further study.

Interestingly, different features of folding sequences were emphasized
in the MS procedure at different input folding 
temperatures. Sequences that were designed
to be stable at high $T$ featured low energy in the native state
and higher dispersion of interaction energies $D$. In contrast,
sequences that were designed to fold at lower temperature
had lower $D$ and higher $E_N$ (see Fig.11 of \cite{MMES}). 
This result shows that an optimal design strategy may be different
for design of thermostable and mesophile sequences. A possible 
reason for that
was discussed in \cite{MMES}.

\section{Designing  longer sequences that fold cooperatively.}

The theoretical approaches to protein design were based on 
the results of mean-field heteropolymer theory which did
not take into account inhomogeneity in the
distribution of interacting aminoacids over 
the protein structure. This approximation neglects the fact
that some parts of the protein, e.g. interior
may have been stabilized to a greater extent than
other parts, e.g. exterior. Lattice simulation showed
that this factor may be important for longer proteins 
giving rise to a ''multidomain'' behavior where core folds
at higher temperature than the surrounding loops, leading
to lower folding cooperativity \cite{DOMAIN,RFIM,FOLDONS}. 
It was shown \cite{DOMAIN,DELTA,FOLDONS} 
that existence of domains
is correlated with $\delta$, the 
dispersion of {\em native} contact energies.
Sequences having higher $\delta$ 
tend to fold less cooperatively
(core first, then loops) while 
sequences with lower $\delta$ fold
as a one cooperative unit. An improved design procedure 
which optimizes both $Z$-score and $\delta$ was proposed
in \cite{DELTA}. This approach  makes it possible 
to design sequences having desired
 folding cooperativity.

\section{Evolution-like design of fast-folding sequences}

Thermal stability is not the only feature of protein sequences that
could be optimized. Another important characteristic is
 folding rate. It is of great interest to compare the sequences 
optimized for stability with the ones optimized for folding rates
because it may shed some light on the features of
proteins that were optimized in natural evolution of their sequences.
The evolution-like selection of fast-folding sequences was
suggested in \cite{EVOL1} and further developed in \cite{EVOL2}. 
The idea of the method is conceptually simple
and similar to the design that optimizes the stability. Mutations
are attempted and only those are accepted that make folding faster
(details are in \cite{EVOL1,EVOL2}). The algorithm has proven successful
yielding many fast-folding sequences. Analysis of the  ''database''
of emerged sequences showed that they are indeed more thermodynamically stable in their
native conformations, than random sequences.   
Interestingly, the $Z$-scores of evolved fast folding sequences
were markedly lower than for random sequences but markedly higher than for sequences 
that were designed to optimize their $Z$-score (we remind
the reader that $Z$ scores are always negative, i.e. 
''lower'' means ''better'', as far as stability is concerned).   
Despite of higher $Z$-score,
sequences generated by evolution-like 
selection procedure
folded much faster than sequences designed
for higher stability (an order of magnitude 
at the respective temperatures 
of fastest folding). 
This points out clearly to the usefulness and 
limitation
of the $Z$-score as predictor of 
the folding rate
(as well as any other
global thermodynamic  criterion).

A more detailed analysis of the features of
evolved fast-folding sequences showed that their 
stabilizing interactions were 
distributed unevenly: acceleration of folding was accompanied
by stabilization of specific fragment of the structure
(the ''folding nucleus'' \cite{NUCLEUS,FER_NUCL,FER_PNAS95,ALAMUT,FER_COSB,COSB}),
while the remaining part of the structure was much less stabilized.
In other words, in the evolution-like selection of fast-folding
sequences the first few mutations lead to the decrease of 
$Z$-score accompanied by some acceleration of folding. 
Further acceleration was achieved after a few 
subsequent mutations that 
strengthened  specific set of contacts,
the folding nucleus. In the steady state
of evolution-like selection where folding rate
did not change much with mutations the
aminoacids at the nucleus positions were
remarkably conserved in contrast to other
positions where mutations were frequent.

A similar approach was taken by 
Nadler and coworkers in their interesting 
study 
of 2-dimensional protein model \cite{NADLER}. 
These authors pointed out
that in their model the energy 
optimization does not always give the desired
results
and additional optimization of folding 
rate may be required to find folding sequences.
This conclusion is consistent with the theoretical 
views presented
in this review (see e.g. Lesson 4): 
Two-dimensional models behave very differently and the
results obtained with these models 
cannot be directly compared with the
results from  three-dimensional models. 
To understand better the differences between two-dimensional
models and three-dimensional ones it is of clear interest to 
study the features of sequences selected for
fast folding in \cite{NADLER}

\section{Lessons for folding}

The best and most objective criterion of
success in protein design is folding of designed
sequences, {\em in vitro}, or {\em in vivo} or 
{\em in silica}. Clearly, certain features of 
the folding phenomenology 
depends crucially on how the sequences were designed/selected.
This fact calls for great caution in comparing folding in
different models where sequences were designed (selected) 
using 
different methods.
In particular sequences that have   large energy gap $E_N-E_c$ fold
cooperatively  (''first order like''). 
In contrast, weakly designed or random heteropolymers 
that do not have such a large gap,
have non-cooperative folding transition. 
\cite{BIOCH,MKES,GROS_BP}). 
Another examples show that such features as on \cite{DOMAIN}
 and off-pathway \cite{TRAPS,FD1} intermediates may be designed ''in''
or ''out'' by proper sequence selection. 

E.g. the folding dynamics for two sequences
designed to fold into the same 36-mer conformation
but using different design strategies were compared
in \cite{FD1} 
The first sequence, $Seq1$
was designed by optimizing the $Z$-score 
(at a variable aminoacid composition) 
while the second one, $Seq2$
was generated using the original 
approach \cite{PNAS}
that minimizes the native state energy 
at constant aminoacid composition.
It was shown that the sequence $Seq1$ 
that was obtained by optimizing
the $Z$-score folded fast, more cooperatively and
was more stable in the native state than $Seq2$. While
the transition for $Seq1$ followed the two-state scenario both
in thermodynamics and kinetics, an 
equilibrium intermediate  and 
structurally similar to
it 
trapped kinetic intermediate
were found  for $Seq2$.

Since both thermodynamics
and kinetics are derived from the properties
of the energy landscape 
there is an established relation between them 
(see e.g. \cite{PHYSKIN}). 
To this end care should be taken in comparing the results
of folding simulations 
for different models in which sequences were designed differently.
Such comparison is possible only if equilibrium
behavior of two models are similar. E.g. recent studies
\cite{GB} showed that folding transition in some off-lattice
models is non-cooperative in contrast to lattice models
and experiment \cite{PRLF,SO3,Privalov96}. This fact 
rules out the nucleation mechanism 
for the model of Ref.\cite{GT}. Correspondingly
it may be not very insightful to compare the cooperative 
kinetics 
 of real proteins and lattice model proteins
with the non-cooperative kinetics in the off-lattice model 
studied in \cite{GT,GB,GT_FD}.

The theoretical developments in protein design 
stimulated interesting
experimental studies including design with reduced, or simplified alphabets
to address the issue of a ''minimalistic'' protein sequence,i.e. 
what is the minimal number of amino acid types that
make it possible to design stable folding sequences. 
Hecht and coworkers \cite{HECHT} 
designed and synthesized sequences based on the 
''two-aminoacid type'' assumption 
that distribution of hydrophobic aminoacids
is most crucial determinant of the structure. While thus designed
proteins were compact and belonged to the expected (helical) 
secondary structure class, their folding into unique structure and cooperativity
has not been fully established. In a recent elegant study by 
Baker and coworkers \cite{Baker_NSB} the phage display 
technique was employed to seek ''minimalistic'' sequences
that fold into the structure of a small protein, SH3, as judged by
its activity. The authors of \cite{Baker_NSB} come to the conclusion
that 6 aminoacids alphabet is generally sufficient for
protein design, with an important exception of a few sites
where simplification was not possible. One possibility
is that these sites are related to function, another
possibility that they participate in the unique folding nucleus.
Future studies will clarify this important issue.

\section{Concluding remarks}

One of the main points of this review is that 
better understanding of protein folding
(at least in the realm of simple models) 
is of crucial importance to the success
of protein design. 

The results of statistical-mechanical analysis 
(see eqs(\ref{eq:s_seq},\ref{eq:REM},\ref{eq:G_max})
and Lesson 1) show
that for an appropriate  model (for which $m_{eff} >\gamma$)
 exponentially (in chain length $N$) large number
of sequences, can fold cooperatively into a given structure.
This is consistent with  the observation that
 many non-homologous
protein sequences can fold into similar conformations
\cite{FSSP}, the fact that  makes 
the ''bioinformatics'' approach to
prediction of protein conformation so difficult.
From the design perspective, the chance that designed sequence
is  identical or even homologous to the native
sequence is minimal. Therefore the success of design
cannot be measured by relatedness of ''predicted'' and native 
primary structure \cite{KD}. However, 
when aminoacids are categorized
into small number of classes, the simplest 
division being into hydrophobic and polar
the correlation between ''predicted'' and 
real sequences is beyond
the noise level \cite{PE}. However, as was 
noted earlier the 
models that have only two kinds of aminoacids   
essentially fail to fold (unless the 
ensemble of conformations is very restricted).

It is almost tautological to say 
that  design represents a search in sequence space
to optimize folding and stability.  The straightforward 
approaches 
 this problem that directly (from simulations) 
evaluate the impact of each mutation on folding
thermodynamics \cite{SENO} or kinetics \cite{EVOL1,NADLER}, are
computationally very intensive and at that point 
are hardly feasible
for models other than simplest lattice model.
This calls for a powerful folding criterion that is 
easy to evaluate without running simulations
in conformational space
after each mutation. Such criterion
should be a good predictor of folding 
ability that can be used  as a ''scoring function'' to be optimized
in sequence space. Here the theory of folding
provides a crucial contribution to design pointing
out to such criteria as energy gap and related to
it $Z$-score as well as $\delta$, the dispersion of energies
of native contacts and in some cases the stability
of the nucleus. Importantly those criteria
are correlated to stability and folding rate (in a certain range
of temperatures, see \cite{MMES,GLASS}) and therefore they
proved very useful for design. A useful folding criterion 
should be simple and easy to evaluate without 
intensive searches in conformational space.
E.g. recently,
the so-called $\sigma$-criterion was proposed
to distinguish between fast- and slow folding sequences \cite{KT}.
While in essence this criterion is related to 
the $Z$-score, or gap criterion 
(\cite{COSB}, A.Dinner,M.Karplus and ES, to be published) its value 
is not known without the folding simulations. That makes the
utility of the $\sigma$-criterion for protein design problematic.

Obviously the folding criteria  that are currently
used for design  have their limitations.
 In particular there is
evidence that fast folding could have 
been an important factor
in evolutionary selection of proteins 
\cite{LIF,EVOL2}. This may call for a criterion
that takes the 
folding kinetics into account more consistently
(a step in this direction was outlined in \cite{CO_PANDE}).  
It is likely that search for better simple folding criteria 
will remain  an important area of research at the interface 
between protein folding and design.

Another crucial bottleneck in protein design is the
lack of knowledge of a potential
function that faithfully reproduces protein energetics
(i.e. for which the native structure for the native sequence
is global energy minimum with energy gap). This direction of research 
has been extremely active (see e.g. in \cite{JB,JT,GSW,MS})
and is likely to be very active in future. The major issue here
is to find a model that is still feasible to simulate
but which has enough detail to make it possible
to derive ''good'' folding potentials. It was shown
in \cite{MS,VD} that simple pairwise contact potential approximation
is too crude to describe real proteins. 
There is no set of parameters U that provides
energy gap that is sufficient 
for successful folding 
simulations of real proteins, 
in the two-body contact approximations
of the energetics. It is almost 
certain that future studies will
seek better potentials for more 
refined models (see e.g. \cite{JB,SMOG1})
that can be used for reliable design
approaches.

A crucial direction of the further study is  
to bring the  progress in theoretical protein design
closer to experiment.  An important issue that needs to be addressed
in applying
theoretical models to the design of real proteins
is whether 
the details of side-chain packing are crucial determinants
of a protein structure. While some original proposals
gave affirmative answer to this question \cite{PR,LS} 
more recent experimental studies indicated that chain flexibility
needs to be taken into account so that many side-chains
substitutions can be accommodated by slightly varying the
backbone conformations \cite{FLEX1,FLEX2}. Interesting methods to
account for side-chain stereochemistry  
in sequence selections have been developed
\cite{MAYO_PNAS,MAYO_SCI,DEAD_END} that use dead-end elimination theorem
or Monte-Carlo design that takes into account 
side chains degrees of freedom \cite{HR}.

An important signature of  the maturity of a field is the degree of interaction
between theory and experiments. By that criterion protein design
enters its maturity stage and we are entitled to witness stunning
progress in the near future.

\section*{Acknowledgments}

I am grateful to Alex Gutin, Victor Abkevich, Leo Mirny and
Mike Morrissey for their
numerous contributions and insights. This work was supported by
NIH grant GM52126

\pagebreak
\section*{Figure Legends}
{\bf 1} A schematic presentation of the protein design problem
(taken from \cite{MMES}):
given the target 3D structure and the selected temperature 
find a sequence that folds at this temperature into
the given conformation and is stable in this conformation.

{\bf 2} Degeneracy of the protein code. The solid line is the analytical
formula (\ref{eq:s_seq}). The average $E_{av}$ and dispersion $D$ 
was calculated
as explained in the text using the 
Myazawa-Jernigan set of parameters (table VI \cite{MJ}). Simulations
using  other parameter set \cite{KGS} provided identical results. 

Data points correspond to the
direct calculation of sequence entropy from MC 
simulations in a range of selective temperatures
(keeping the aminoacid composition same 
as in the native sequence). 
Average energy of sequences in the 
target structure $E(T)$ was evaluated from
simulation runs
Then eq.(\ref{eq:S_T}) was applied to 
obtain sequence space entropy. (Here we show
 entropy and energy , normalized per aminoacid 
residue $s_{seq}=S_{seq}/N$, $e_N=E_N/N$). 
Different symbols correspond to different proteins
(in pdb access code): filled diamonds - 4mbn, 
open squares - 2cab, filled squares - 1pcy,
open diamonds - 2pal. Horizontal insert 
is given for illustrative purpose 
to show schematically the generic representation 
of density
of states in conformational space, as predicted by the heteropolymer
theory \cite{BIOCH,GROS_BP}. 
The range of energies at which density of non-native
decoys is high is shown in black, a few low 
energy conformations (shown as discrete lines
in the insert) that lie below the
boundary of continuous spectrum $E_c$ represent lowest energy decoys. 

{\em a)} ''Designable model'' where $m_{eff} > \gamma$. 
Many sequences ($ \sim exp(1.9N)$ in the present
example)  
exist that have low energy $E_N$ in the target 
conformation with pronounced stability
gap $\Delta=E_N-E_c$. Such sequences are expected to fold fast into
the native conformation

{\em b)}
Non-designable model $m_{eff} < \gamma$: no sequences that fold uniquely 
to the ground state can be found. The model
runs out of sequences at energies which are not low enough to 
ensure large gap  between the native structure
and misfolded decoys. The data points represent MC design simulations
entropy for two ''HP'' models of 
proteins: 1mbn (upper curve) and 1pcy (lower curve). 
Aminoacids
were categorized into ''H'' and ''P'' classes 
as explained in \cite{PE}. The more pronounced
difference between proteins is due to the 
difference in their average hydrophobicities. 
i.e fraction of hydrophobic residues
in their sequences.

{\bf 3} The density of states (energy spectra) 
in the ensemble of  fully compact conformations
of the 27-mer model for
a random and best designed sequence.
Each bar corresponds to entropy per residue - 
the logarithm of the number of all conformations
having given energy divided by the number of residues (27 in this
case).
The density of states plots are 
derived from exhaustive enumeration 
of all 103346 compact conformations of the 27-mer \cite{NATUR}.
For simplicity  only two types of monomers
are used (''black'' and ''white'') with nearest neighbor ''color specific''
interactions: $E_{BB}=E_{WW}=-3$; $E_{BW}=-1$ \cite{PNAS,SO2}. 
While this interaction matrix may be not quite realistic
for real proteins, it is useful for clarifying basic concepts
presented in this review.
Obviously the
lowest energy conformation is the one that maximizes the number of 
favorable ''same color'' (SC) contacts. Left insert shows the target
structure and the sequence that has minimal possible energy
$E_{lowest}= -84$ (all 28 contacts are SC) in that structure. 
This structure
represents a unique ground state for
the designed sequence: The black bar for the designed
sequence corresponding
to the energy $E_N=-84$  is slightly
exaggerated to make it visible.
  The right insert shows the same
structure with a quasirandom sequence fit into it.

\pagebreak

\pagebreak
\pagestyle{empty}

\begin{center}
\par \leavevmode\def\epsfsize#1#2{0.8 #1}\epsffile{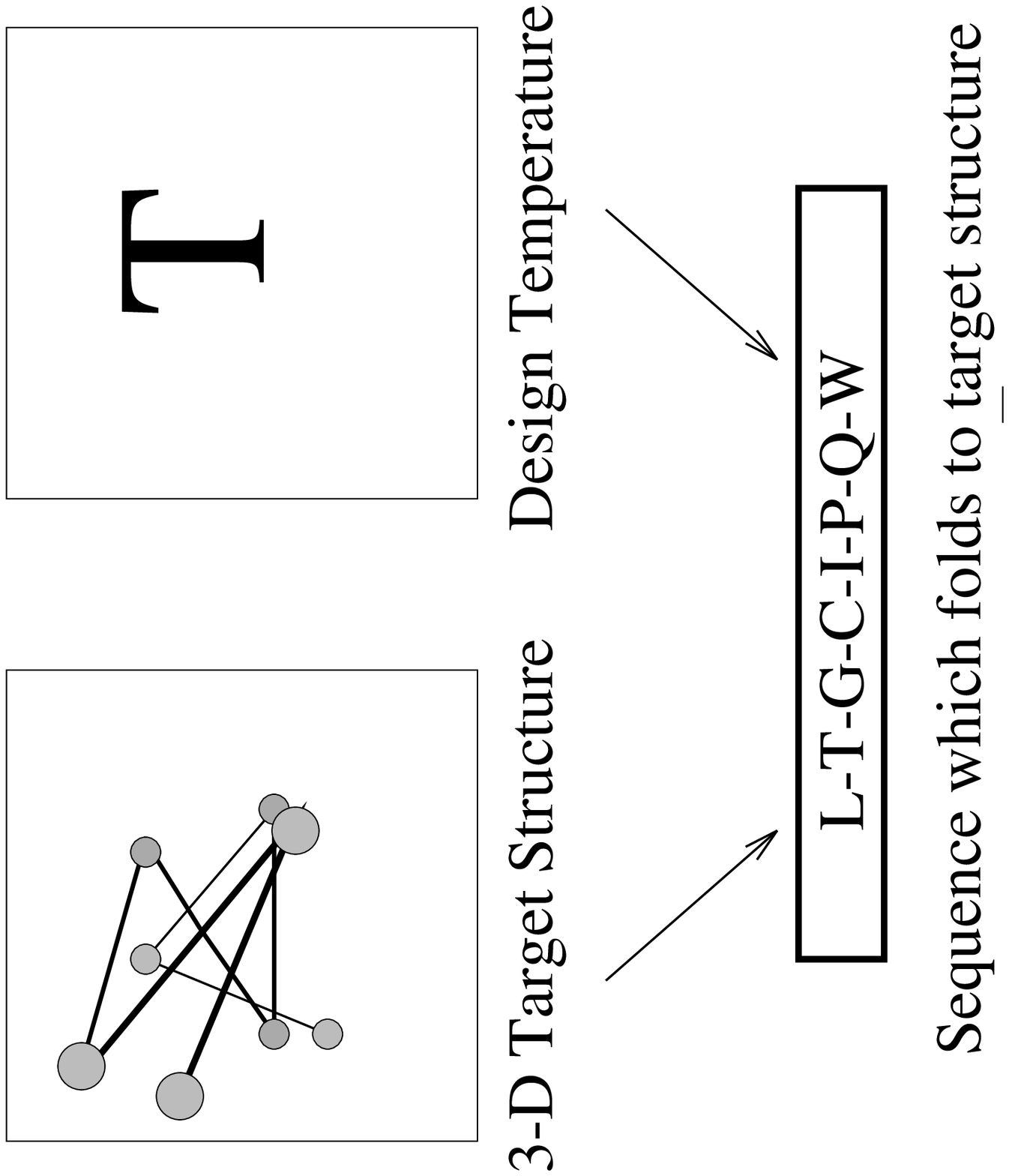}

\begin{flushright}
''Protein design..''. Fig.1
\end{flushright}
\end{center}

\pagebreak

\begin{center}
\par \leavevmode\def\epsfsize#1#2{0.8 #1}\epsffile{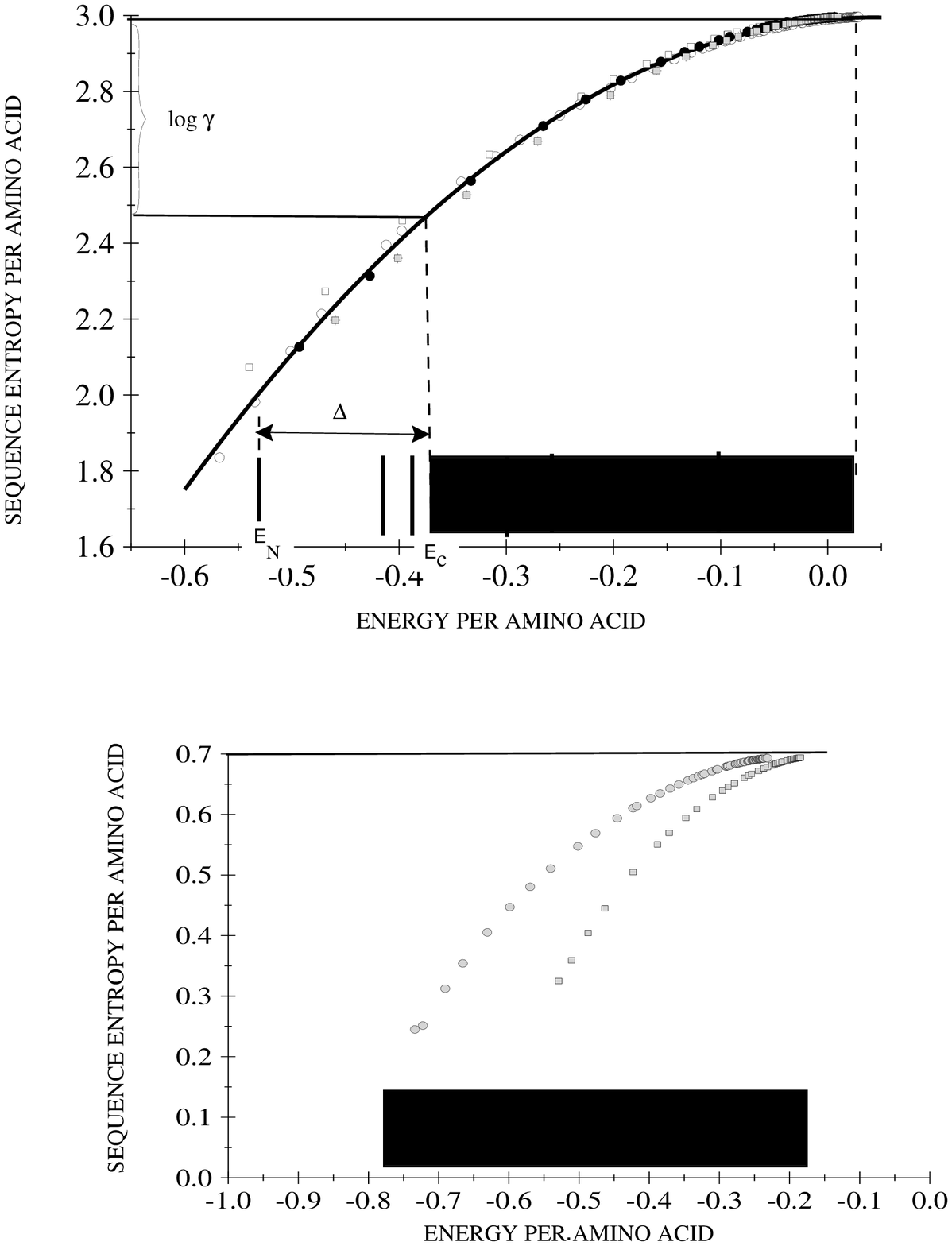}

\begin{flushright}
''Protein design..''. Fig.2
\end{flushright}
\end{center}

\pagebreak

\begin{center}
\par \leavevmode\def\epsfsize#1#2{0.8 #1}\epsffile{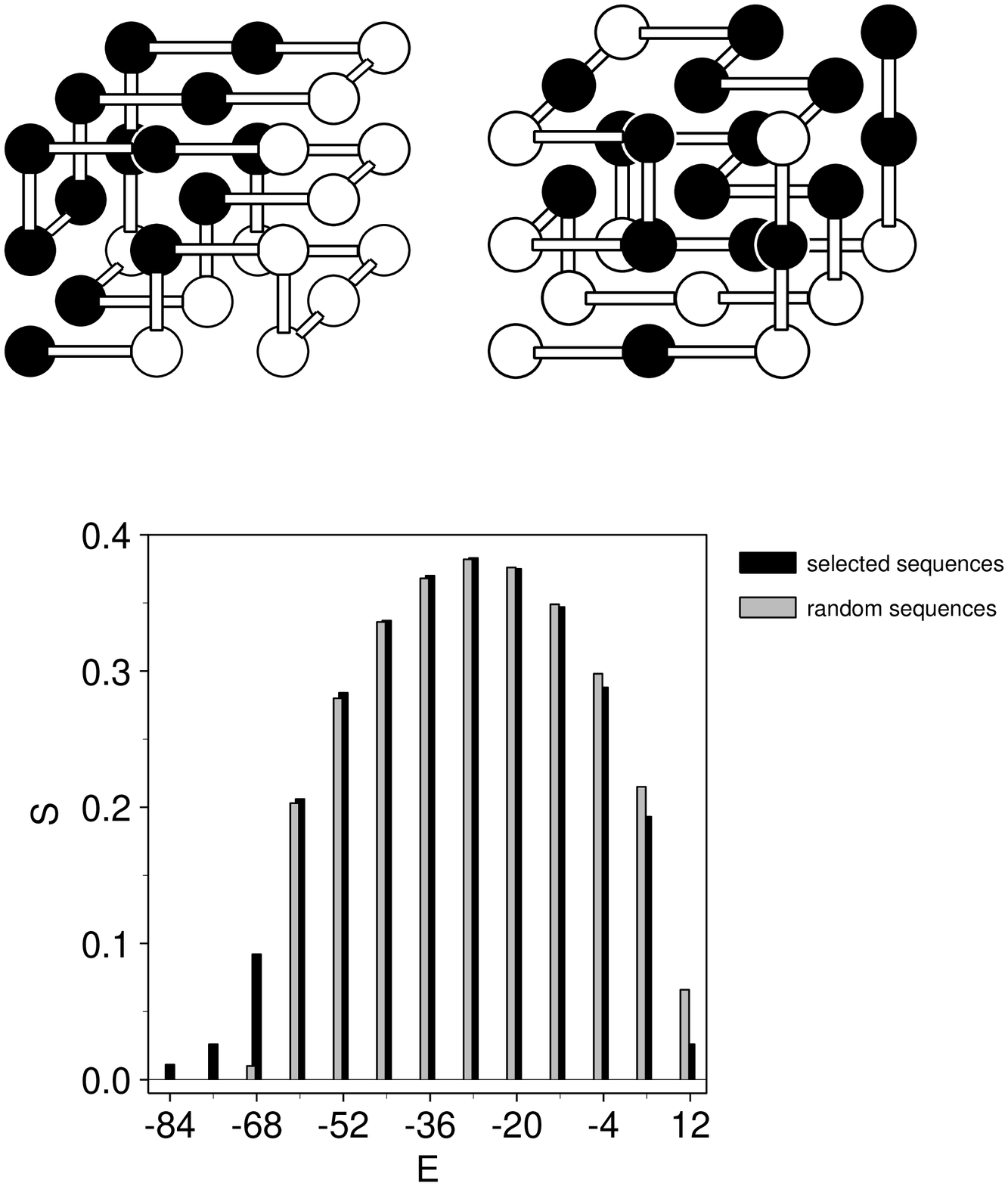}

\begin{flushright}
''Protein design..''. Fig.3
\end{flushright}
\end{center}

\end{document}